\documentstyle[preprint,aps]{revtex}
\title{Comment on ``Evolution from BCS Superconductivity to
Bose-Einstein Condensation: Role of the Parameter $k_{\rm F} \xi$ in
Interpreting the Experimental Plot by Uemura et al.''}
\author{J.J. Rodr\'{\i}guez-N\'u\~nez$^{1,3}$\footnote{Permanent
Address:
Departamento de F\'{\i}sica, Grupo de S\'olidos, FEC - LUZ,
Apartado 526, Maracaibo, Venezuela.  Also a Visiting
Scientist at IVIC, Centro de F\'{\i}sica, Apartado 21827,
Caracas 1020-A, Venezuela,
\newline
Corresponding author:
J.J. Rodr\'{\i}guez-N\'u\~nez, E-mail: odr@zurich.ibm.com},
S. Schafroth$^2$,
T. Schneider$^1$, M.H. Pedersen$^{1,2}$ and C. Rossel$^1$}
\address{$^1$IBM Research Division, Zurich Research Laboratory,
CH-8803 R\"uschlikon, Switzerland $^2$Physik-Institut der
Universit\"at Z\"urich, Winterthurerstrasse 190,
CH-8057 Zurich, Switzerland
$^3$Institut de Physique,
Universit\'e de
Neuch\^atel,
Rue A.L. Breguet 1, CH-2000 Neuch\^atel,
Switzerland}
\begin{document}
\draft
\maketitle
\begin{abstract}
        The problem of crossover from BCS to Bose-Einstein
condensation has recently attracted considerable interest
owing to the discovery of high-$T_{\rm c}$ cuprates. Their
short coherence length, $\xi$, places these materials
in the
interesting region between BCS and Bose-Einstein condensation. In
the paper of F. Pistolesi and G.C. Strinati (Phys. Rev. B {\bf 49},
6356 (1994)), the Nozi\`eres and Schmitt-Rink approach
(NSR) is taken,
which is valid for the weak coupling regime. They derive a relation
between $T_{\rm c}$ and $T_{\rm F}$, which they
insist is valid for any
coupling strength. We present arguments that their assumptions are
incorrect by using our fully self-consistent T-matrix
formalism in two dimensions, and
show that the NSR approach produces unphysical results in this
case.
\end{abstract}
\clearpage
    In a recent paper, Pistolesi and Strinati\cite{Pistolesi}
attempt to justify a
variable that is able experimentally to describe the transition from
extended pairs (BCS approach) to local pairs (Bose condensation). They
do so because they feel that there must be a better variable than
the interaction parameter, $-V$. Such a variable
should be independent of the
superconducting pairing mechanism. In the end,
they conclude that
such a variable should be $k_{\rm F} \xi$, where $k_{\rm F}$ is the
Fermi
$k$-vector and $\xi$ is the coherence length. With this variable
they locate the high-$T_{\rm c}$ superconductors close to the
instability
$k_{\rm F} \xi \approx 2\pi$ in the plot of $T_{\rm c}$ vs.
$T_{\rm F}$, which they call the Uemura plot\cite{plot}.

Let us preface
our discussion of the details of
Pistolesi and Strinati's paper with the statement that the
important quantities to plot are $T_{\rm c}$ vs. ${n_{\rm s}}/{m}$,
where
$n_{\rm s}$ is the superfluid density and $m$ is the effective mass
of the
pairs. The value of
$T_{\rm c}$ vs.\ ${n_{\rm s}}/{m}$ increases in
the underdoped regime, saturates for maximum doping and then decreases
in the overdoped regime. This is not discussed in the paper of
Pistolesi and Strinati. For example,
$\varepsilon_{\rm F}$ and $T_{\rm F}$ are quantities that increase
with doping.
Therefore, the variable $\varepsilon_{\rm F}$ is not the correct
scaling variable
because it cannot explain the variation of $T_{\rm c}$ vs.\
$\sigma(0)$, where
$\sigma(0)$ is the zero temperature muon relaxation rate. To be more
specific, plotting $T_{\rm c}$ vs.\ doping produces
different curves for different materials. However, when
normalizing $T_{\rm c}$ by the maximum transition temperature, all the
curves collapse on a generic curve\cite{Toni-Hugo}.
This cannot be explained with the
standard BCS approach.

In
addition, recent experiments with high-$T_{\rm c}$ superconductors also
reveal
features that cannot be explained by the standard BCS
approximation. For example, upon doping, the materials go from an
insulating to a superconducting phase, where $T_{\rm c}$
increases until a maximum doping of about 0.16
electrons/CuO\cite{Tallon} is reached.
As doping continues,
$T_{\rm c}$ decreases again and a
transition to a metallic state occurs around
0.27 electrons/CuO\cite{Tallon}.
Hence, the
simple BCS
description of the problem is not satisfactory, whereas a Bose
scenario appears more appropriate, as has been found by
Schneider and Pedersen\cite{Toni-et-al}, among others.

We shall now raise a
few technical points regarding
the arguments of
Pistolesi and Strinati for using Eq.\ (9) in their paper.

\begin{enumerate}
\item
They do not take the discreteness of the lattice into account,
even
when the high-$T_{\rm c}$
superconductors are strongly correlated electron
systems and the band plays an important role.
The continuum approximation is
valid only in the dilute limit. The assumption of a parabolic band is
clearly not valid for all dopings.
The reason they retain the
variable $k_{\rm F}$ is that they have not considered the
full band structure.

\item
They use the Nozi\`eres and Schmitt-Rink\cite{NSR} approach (NSR)
to omit the interaction potential in favor of $k_{\rm F} \xi$.
However, $k_{\rm F}$ is clearly a variable that is only meaningful
for a
weakly interacting, many-particle Fermi system. For example, with
the two-dimensional attractive Hubbard model\cite{ours}, which adopts
the T-matrix
approach, we have found that the distribution
function, $n(k)$, differs from the the Fermi
distribution function by  $20 \%$ when $U/t = -4.0$ and $n = 0.32$,
where $t$
is the hopping matrix element between nearest neighbors, $U$
the on-site
attraction between electrons with opposite spins and $n$ the total
number of carriers. In this case, therefore,
$k_{\rm F}$ is not defined exactly,
and the best we can do is to define it
as the point where $n(k)$ reaches half its maximum value.

\item
The BCS trial wave function is robust at zero
temperature, where fluctuations are irrelevant. But at higher
temperatures, it is better
to use the Thouless criterion to evaluate $T_{\rm c}$
with the T-matrix approach. The coherence length,
$\xi$, should be calculated from the correlation function at
large distances. Hence, the use of Eq. (9) in Pistolesi and
Strinati's paper is not correct. More appropriate would be
two-particle
correlation functions, which in the language of the T-matrix approach
is the T-matrix function itself.

\item
The NSR scheme fails for strong couplings for which the
approach of Alvarez and Balseiro\cite{Balseiro}
has proved to be a good one.
As a consequence, we conclude that there is no unique approach
that allows us to go from the weak coupling regime to the strong
coupling one. On the other hand,
the NSR scheme has the questionable feature that for all
densities, in 2D\cite{two}, the ground state is defined by bound pairs
for any value of the interaction $U$.

We strongly
emphasize here that the T-matrix approach is not valid down to
zero temperature, unless we generalize it to include the anomalous
Green function.
In fact, according to our calculations (Fig.\ \ref{fig1})
with the fully self-consistent
T-matrix approximation, we obtain the critical
temperature, $T_{\rm c}$, for a particular density, and
$T_{\rm c} \neq 0 $.
It is determined using
the divergence of ${\rm Re}(T({\vec 0},0))$
or the Thouless criterion. ${\rm Re} (...)$ denotes the
real part of what follows.
For $T < T_{\rm c}$, we cannot use the T-matrix approach in its simple
form,
because the
instability breaks the symmetry.
Consequently, we should not use the T-matrix formalism in 2-D
in the way
proposed by Nozi\`eres and Schmitt-Rink, because the
temperature cannot be lowered
to approach
the ground state. It produces unphysical results
like those obtained by Schmitt-Rink et al.\ \cite{two}, see
Fig.\ 3 in \cite{two} in particular.
For this reason, Nozi\`eres and Schmitt-Rink's approach
must be used with care.

\item
Pistolesi and Strinate
plot $\mu$ vs.\ $k_{\rm F} \varepsilon$ and normalize
$\mu$ differently for $\mu > 0$ and $\mu < 0$. There is no
reason
{\it a priori}
why $\mu$ should interpolate smoothly in the entire range of
$k_{\rm F} \xi$. In fact, it does not. They take the value $k_{\rm F}
\xi \approx 2\pi$ as the criterion for an instability towards
Bose condensation. From this condition, they conclude that $\xi =
\lambda_{\rm F}$, where $\lambda_{\rm F}$ is the Fermi wavelength.
However, the
important question to ask is why this instability occurs at $\xi =
\lambda_{\rm F}$,
which they do not explain.

\item
High-$T_{\rm c}$ materials are highly anisotropic, which is
reflected by such quantities as the coherence length parallel
and perpendicular to the copper-oxide
planes, i.e., $\xi_{||}$ and $\xi_{\perp}$,
respectively. This is not discussed in Pistolesi and Strinati's
paper either. For example, Schneider et al.\cite{Toni3} find
that the behavior of $T_{\rm c}$ vs.\ $\rho$,
where $\rho$ is the band filling, depends on the parameter
$\alpha$, defined by $\alpha = {t^2_{\perp}}/{t^2_{||}}$, where
$t_{\perp}$ and $t_{||}$ are
the hopping integrals perpendicular to and
on the planes.
\end{enumerate}

   In conclusion, the approach taken by Pistolesi and Strinati
lacks of physical meaning and does not take recent
measurements\cite{Uemura} of
$T_{\rm c}$ vs.\ $\sigma(0)$ into account, where
$T_{\rm c}$ is observed to increase and then to
decrease with doping. The term
$\sigma(0)$
is the muon spin relaxation rate at zero temperature.
Pistolesi and Strinati have used the parabolic band approximation,
but it is the very fact that the bands are narrow that enables
the high-$T_{\rm c}$ superconductors to have universal
properties\cite{Toni3}. In Ref.\ \cite{Toni3}, $T_{\rm c}$ is found
to increase with increasing carrier
concentration, reach a maximum and then come down to zero. See,
for example, Fig.\ 2 in \cite{Toni3}. The
universal properties of high-$T_{\rm c}$ superconductors must
therefore be studied
using a narrow band description.
Furthermore, above $T_{\rm c}$, the description of the high-$T_{\rm c}$
materials
must include pair
fluctuations, and the critical temperature
should be evaluated from the fully self-consistent
T-matrix formalism by Thouless criterion. The value of
$T_{\rm c}$
should not be calculated from the BCS gap
equation, i.e., from a mean field approximation. We have used the
fully self-consistent T-matrix formalism above $T_{\rm c}$ and
found that the approach followed by Nozi\`eres and
Schmitt-Rink produces unphysical results. In addition,
we have calculated the distribution function\cite{ours} in the
strongly correlated limit ($U/t~=~-4.0$) and found that the
Fermi energy is not a well-defined quantity.
As the approach taken
by Pistolesi and Strinati does not have a real physical
basis, their conclusion is not justified.

   We would like to thank H. Beck, J.-P.\ Locquet and Ariel
Dobry for very interesting
discussions. We gratefully acknowledge the support of the Swiss
National Science Foundation (NFP 30).
One of the authors (JJRN) acknowledges
partial
support both from {\bf CONICIT}, through project {\bf No. F-139}, and
{\bf CONDES-LUZ}.
We would like to thank
Mar\'{\i}a Dolores Garc\'{\i}a-Gonz\'alez for reading the manuscript.

\begin{figure}
\caption{
The real part of the T-matrix at zero momentum and zero
frequency as a function of temperature. The number of points in the
Brillouin zone is $22 \times 22$
and the number of the Matsubara frequency is
1024. We have chosen a damping factor of $\delta = 0.1$ to go to real
frequencies.}
\label{fig1}
\end{figure}
\end{document}